\documentclass[aps,prd,twocolumn,showpacs,nofootinbib,amsmath,amssymb,floatfix,superscriptaddress,showkeys]{revtex4}
\usepackage{graphicx}
\usepackage{dcolumn}
\usepackage{bm}
\usepackage{captcont}
\usepackage{xcolor}
\usepackage{threeparttable}

\usepackage{epstopdf}
\usepackage{mathtools}
\usepackage{natbib}
\usepackage{ulem}
\usepackage{color,txfonts}
\definecolor{blue0}{rgb}{0,0,0.6}
\usepackage[colorlinks,linkcolor=blue0,anchorcolor=blue0,citecolor=blue0,urlcolor=blue0]{hyperref}

\usepackage{color,amsmath,amssymb,graphicx,latexsym,lineno}
\usepackage{threeparttable,multirow,txfonts,subfigure,booktabs}
\newcommand{\jcap}{J. Cosmol. Astropart. Phys.}

\newcommand{\mnras}{Mon. Not. R. Astron. Soc.}
\newcommand{\araa}{Annu. Rev. Astron. Astrophys.}
\newcommand{\aap}{Astron. Astrophys}
\newcommand{\aj}{Astron. J.}

\newcommand\pasa{Publ. Astron. Soc. Australia}

\begin{document}
\title{Constraining ultralight dark matter using the Fermi-LAT pulsar timing array}
\author{Zi-Qing Xia}
\affiliation{Key Laboratory of DM and Space Astronomy, Purple Mountain Observatory, Chinese Academy of Sciences, Nanjing 210033, China }
\author{Tian-Peng Tang}
\affiliation{Key Laboratory of DM and Space Astronomy, Purple Mountain Observatory, Chinese Academy of Sciences, Nanjing 210033, China }
\affiliation{School of Astronomy and Space Science, University of Science and Technology of China, Hefei, Anhui 230026, China }
\author{Xiaoyuan Huang\footnote{Corresponding Author: xyhuang@pmo.ac.cn}}
\affiliation{Key Laboratory of DM and Space Astronomy, Purple Mountain Observatory, Chinese Academy of Sciences, Nanjing 210033, China }
\affiliation{School of Astronomy and Space Science, University of Science and Technology of China, Hefei, Anhui 230026, China }
\author{Qiang Yuan\footnote{Corresponding Author: yuanq@pmo.ac.cn}}
\affiliation{Key Laboratory of DM and Space Astronomy, Purple Mountain Observatory, Chinese Academy of Sciences, Nanjing 210033, China }
\affiliation{School of Astronomy and Space Science, University of Science and Technology of China, Hefei, Anhui 230026, China }
\author{Yi-Zhong Fan\footnote{Corresponding Author: yzfan@pmo.ac.cn}}
\affiliation{Key Laboratory of DM and Space Astronomy, Purple Mountain Observatory, Chinese Academy of Sciences, Nanjing 210033, China }
\affiliation{School of Astronomy and Space Science, University of Science and Technology of China, Hefei, Anhui 230026, China }

\date{\today}

\begin{abstract}

Ultralight dark matter (ULDM) is proposed as a theoretical candidate of dark matter particles with masses of approximately $10^{-22}$ eV.
The interactions between ULDM particles and standard model particles would cause variations in pulse arrival times of millisecond pulsars, which means that the pulsar timing array (PTA) can be used to indirectly detect ULDM.
In this letter, we use the gamma-ray PTA composed of 29 millisecond pulsars observed by the Fermi Large Area
Telescope (Fermi-LAT) to test four ULDM effects, including gravitational effects for generalized ULDM with different Spin-0/1, the fifth-force coupling effect of dark photon, and the modified gravitational effect of the Spin-2 ULDM.
The gamma-ray pulsar timing is not affected by the ionized interstellar medium and suffers relatively simple noises, unlike that of the radio band.
Our work is the first time that the gamma-ray PTA has been used to search for the ULDM.
No significant signals of ULDM are found based on the Fermi-LAT PTA for all four kinds of ULDM models. 
Constraints on ULDM parameters are set with the 95\% confidence level, which 
provides a complementary check of the nondetection of ULDM for radio PTAs and direct detection experiments.

\end{abstract}

\maketitle

\section{Introduction}\label{Introduction}
Dark matter (DM) is believed to make up almost a quarter of the total energy of the current Universe, which is more than 5 times of ordinary visible matter~\cite{1502.01589}.
Theoretical physicists have proposed various hypothetical particles as dark matter candidates, such as weakly interacting massive particles (WIMPs)~\cite{WIMP2011}, axions~\cite{axion1978}, sterile neutrinos~\cite{1807.07938} and dark photons~\cite{1311.0029}.  
The masses of these proposed dark matter particles span a broad range from below $10^{-22}$ eV to above TeV. 

Here, we focus on the ultralight dark matter (ULDM) with the mass of $\sim 10^{-22}$ eV, whose de Broglie wavelength is up to the sub-Galactic scale ($\sim$ 1 kpc).
Compared with other dark matter candidates, ULDM have the advantage in forecasting small-scale structures consistent with observations~\cite{0003365,0805.4037,0909.0515,1103.3705,1611.00892,1707.04256, 1908.02508,1912.07064}.
ULDM models can be roughly classified by the spin of the particle: (1) the Spin-0 (scalar or pseudoscalar) ULDM case, like ultralight axion-like scalar field dark matter~\cite{1408.4670, 1610.08297, 1612.06789,1705.04367,Yuan_2021}; (2) the Spin-1 (vector) ULDM case, especially ultralight dark photon (DP)~\cite{Guo_2019,2019PhRvL.123b1102D,2005.01515,2105.04565,PhysRevD.106.103024}; and (3) the Spin-2 (tensor) ULDM case~\cite{Marzola:2017lbt,Aoki:2017cnz,Novikov2016}.
Because of the macroscopic de Broglie wavelength, the gravitational effect of generalized Spin-0/1 ULDM and the coupling effect (usually dubbed as the fifth force) of dark photon with Spin-1 would cause oscillations of Earth and pulsars. 
For the Spin-2 ULDM, photons propagating from pulsars to Earth would be affected by the modified gravitational effect and follow the geodesics of the new metric related to the surrounding Spin-2 ULDM field. 
These effects for all these cases would result in monochromatic periodical variations in the pulse arrival time and provide an interesting method to probe ULDM with the pulsar timing array (PTA).
Though another multi-field ULDM scenario has also been discussed for a wide-band spectrum in Ref.~\cite{2112.15593}, here we mainly focus on the monochromatic signal induced by ULDM in the nanohertz range using the PTA dataset.
In addition to the PTA, Ref.~\cite{2023PhRvL.130k1401C} innovatively proposed utilizing photon ring astrometry from the Event Horizon Telescope to detect the presence of the ULDM.

The PTA experiments have been continuously monitoring a series of highly-stable millisecond pulsars (MSPs) for more than a decade, and accurately recording the arrival times of their periodic electromagnetic pulses. 
Gravitational perturbations could induce variations in the time that pulses take from a millisecond pulsar to Earth.
Therefore, PTA is designed to be an excellent detector for nanohertz-frequency gravitational perturbations from the Gravitational Wave Background (GWB) and the single gravitational wave event, and also for the oscillations induced by ULDM~\cite{0911.5206,1006.3969, 1309.5888, 1810.03227, 1912.10210,2210.03880, 2112.07687, 2005.03731, 2110.03096, 2022arXiv220902741U, 2205.06817}. The traditional PTA projects are generally at the radio wavelength: the Parkes Pulsar Timing Array (PPTA) in Australia \cite{Manchester2013,PPTADR2}, the European Pulsar Timing Array (EPTA)~\cite{Kramer_2013,Desvignes2016}, the North American Nanohertz Observatory for Gravitational Waves (NANOGrav)~\cite{McLaughlin_2013,Arzoumanian_2020} and that for the 500-m aperture spherical radio telescope (FAST)~\cite{FAST2011,2018FAST}.
Recently, a gamma-ray PTA recently has been constructed from the Fermi Large Area Telescope (LAT) data for searching GWB from merging supermassive black hole binaries~\cite{2204.05226}, which can provide an independent and complementary check for radio PTAs. Also, the gamma-ray PTA has the advantage that it would not involve the uncertainty from the effect of the ionized interstellar medium (IISM) on the propagating path of photons which can induce additional noises in radio PTAs.
In this letter, the public timing data of the Fermi-LAT PTA\footnote{\url{https://zenodo.org/record/6374291\#.YzVcbC-KFpR}} is adopted to search for the signals produced by the coherent oscillation effect of ULDM particles with different spin\footnote{After our work appeared on arXiv, another work used the Fermi-LAT PTA data to study spin-0 ULDM appeared recently \cite{2023arXiv230404735N}.}, including the gravitational effects for the Spin-0/1 ULDM, the fifth-force effect of Spin-1 dark photon, and the modified gravitational effect for the Spin-2 ULDM.

In this work, the amplitude of ULDM field is assumed to be proportional to the local density of dark matter ($\rho_{\rm DM}$) or its square root (except for the fifth-force case). Considering that measurement time is much shorter than coherence time ($\tau_c\sim10^6/f$ where $f$ is the frequency of the oscillating ULDM field), the amplitude is random following a distribution rather than the deterministic relation \cite{2018PhRvD..97l3006F,Centers:2019dyn,2022JCAP...06..014C}. 
This {\it stochastic} treatment will result in a change of the final results on e.g., the coupling between ULDM and standard model field, by a factor of $1\sim 10$ compared with the {\it deterministic} treatment \cite{Centers:2019dyn,2022JCAP...06..014C}. 
Here, for the convenience of comparison with previous works~\cite{1810.03227,2210.03880,2005.03731} based on the radio PTA, we keep the {\it deterministic} treatment for most cases, except for the fifth force signal of the dark photon in which the stochastic distribution of the amplitude is taken into account \cite{2112.07687}. The effect due to the stochastic nature of the ULDM field will be explored in detail in future works. It is expected that similar impacts on the radio and gamma-ray PTAs will be obtained when considering the stochastic distribution of the ULDM amplitude in the Bayesian inference.

\section{Ultralight dark matter model}\label{sec:ULDMmodel}

In our work, we focus on four ULDM effects: the gravitational effects for ULDM with Spin-0/1, the fifth-force effect of dark photon and the modified gravitational effect of the Spin-2 ULDM.
The detailed descriptions of signals for these four effects are shown in the Appendix Sec.A.
For the Spin-0 gravitational signal, the ULDM would induce oscillations of Earth and pulsars and cause periodical residuals in times of arrival (TOA) of pulse, which can be written as~\cite{1309.5888,1810.03227}:
\begin{equation}
s(t)=\frac{\Psi}{m}\sin(\theta_{e}-\theta_{p})\cos(2 m t+\theta_{e}+\theta_{p})\, ,
\label{eq:st1-spin0}
\end{equation}
where $m$ is the mass of ULDM, and $\theta_{e}$ and $\theta_{p}$ characterize oscillation phases for Earth and pulsar terms, respectively.
The oscillating amplitude $\Psi$ depends on the density of dark matter $\rho_{\text{DM}}$:
\begin{equation}
\Psi=\frac{G \rho_{\text{DM}}}{\pi f^2}\approx 6.1 \times 10^{-18} \left(\frac{m}{10^{-22}\,\text{eV}}\right)^{-2} \left(\frac{\rho_{\text{DM}}}{\rho_{0}}\right)\, ,
\label{eq:Psi-spin0}
\end{equation}
where $\rho_{0}=0.4 \, \text{GeV cm}^{-3}$ is the measured local dark matter density near Earth~\cite{1205.4033,1708.07836}.
Different from the Spin-0 case, the gravitational signal of the Spin-1 ULDM would have an extra contribution from the traceless spatial metric perturbations, related to the oscillation angle $\theta$ ~\cite{1912.10210,2210.03880}:
\begin{equation}
s(t) = -\frac{3}{8} \left(1+2\cos2\theta\right)\frac{h_{\mathrm{osc}}}{m}\sin(\theta_{e}-\theta_{p})\cos(2 m t+\theta_{e}+\theta_{p})\, ,
\label{eq:st2-spin1}
\end{equation}
where the amplitude $h_{\mathrm{osc}}$ is also dependent on $\rho_{\text{DM}}$:
\begin{equation}
h_{\mathrm{osc}}=\frac{8 \pi G \rho_{\text{DM}}}{3 m^2}\approx 1.7 \times 10^{-17} \left(\frac{m}{10^{-22}\,\text{eV}}\right)^{-2} \left(\frac{\rho_{\text{DM}}}{\rho_{0}}\right)\,.
\label{eq:h-spin1}
\end{equation}
In the case of the ultralight dark photon, its fifth force could also lead to oscillations of Earth and pulsars and induce the TOA residuals.
Here we consider two new gauge interactions ($U(1)_B$ and $U(1)_{B-L}$), as shown in the Appendix Sec.A.3.
The timing residuals caused by the ``fifth-force" effect can be given by~\cite{1801.10161,2112.07687}:
\begin{eqnarray}
\label{eq:stb-spin1}
s(t)^{(B)} &=&  \frac{\epsilon e}{m} \boldsymbol{A_0} \left[ \frac{q^{(B)}_e}{m_e} \cos\left(m t+\boldsymbol{\theta}_e\right)\right. - \nonumber \\
& &\left.\frac{q^{(B)}_p}{m_p} \cos(m t+\boldsymbol{\theta}_p)  \right]\cdot \boldsymbol{n}, \\
\label{eq:stbl-spin1}
s(t)^{(B-L)} &=&  \frac{\epsilon e}{m} \boldsymbol{A_0} \left[ \frac{q^{(B-L)}_e}{m_e} \cos\left(m t+\boldsymbol{\theta}_e\right)\right. - \nonumber \\
& &\left.\frac{q^{(B-L)}_p}{m_p} \cos(m t+\boldsymbol{\theta}_p)  \right]\cdot \boldsymbol{n},
\end{eqnarray}
where $\epsilon$ is the coupling strength of new gauge interaction, $e$ is the electromagnetic coupling constant, $\boldsymbol{A_0}$ is the amplitude of dark photon field, $\boldsymbol{n}$ is the normalized position vector pointing from Earth to the pulsar, $q^{(B)}_{e,p}$, $q^{(B-L)}_{e,p}$, and $m_{e,p}$ each represent the $B$ number, the $B-L$ number, the mass of Earth and the pulsar.
As for the Spin-2 case, its modified gravitational effect could have a significant impact on the propagation of photons from pulsars to Earth and result in the TOA residuals given by~\cite{2005.03731}:
\begin{eqnarray}
\label{eq:st-spin2}
s(t) &=& \frac{\alpha \sqrt{2\rho_{0}}}{\sqrt{15} m^2 {M_{\text{P}}}}\cos\left(mt+\frac{\theta_{e}+\theta_{p}}{2}\right)\, ,
\end{eqnarray}
where $\alpha$ is the dimensionless strength constant and ${M_{\text{P}}}$ is the reduced Planck mass.
Prior distributions of parameters for all four ULDM effects are summarized in Tab.~S1 in the Appendix Sec.A.

\begin{figure*}[!htb]
\centering 
\includegraphics[width=0.495\textwidth]{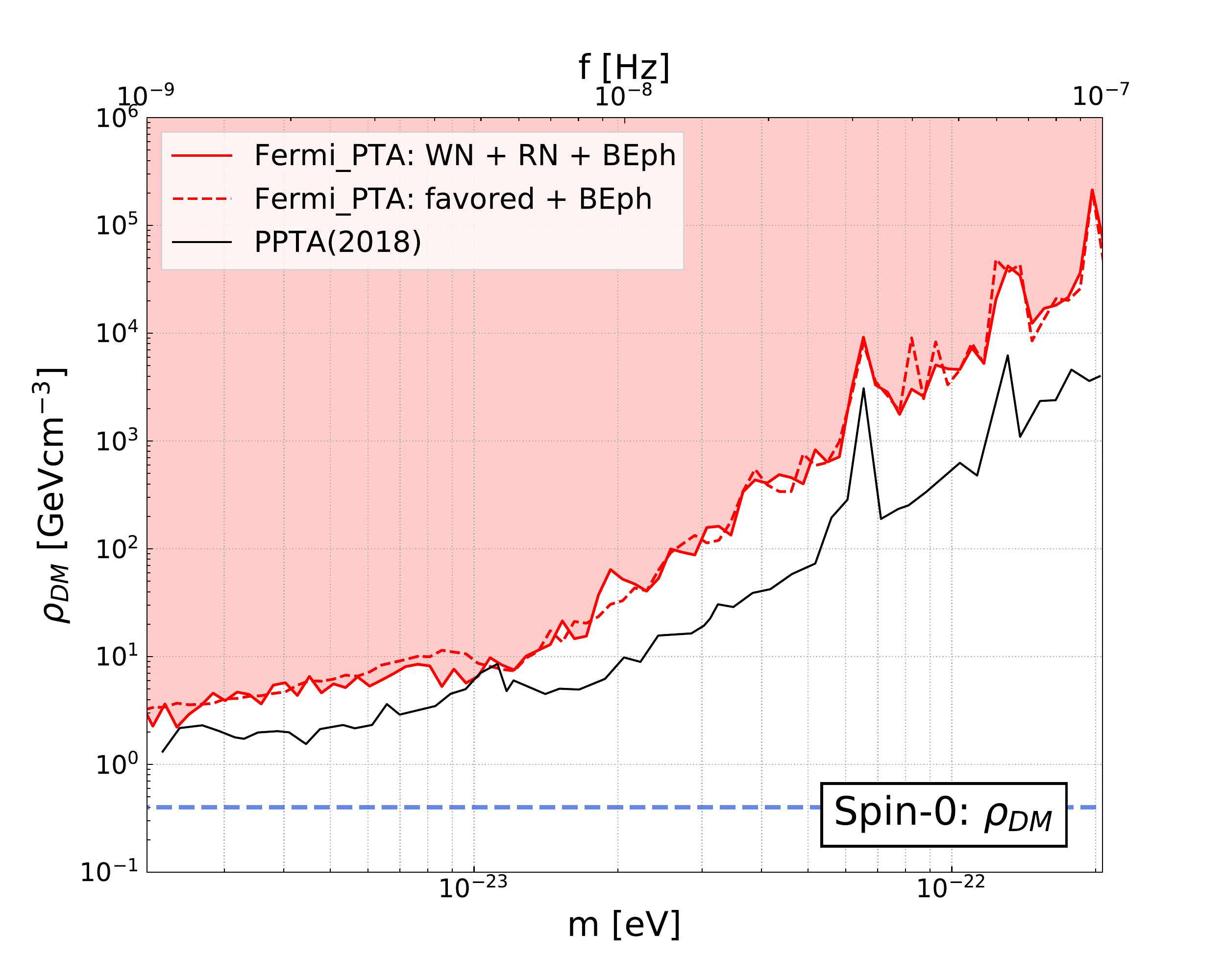}
\includegraphics[width=0.495\textwidth]{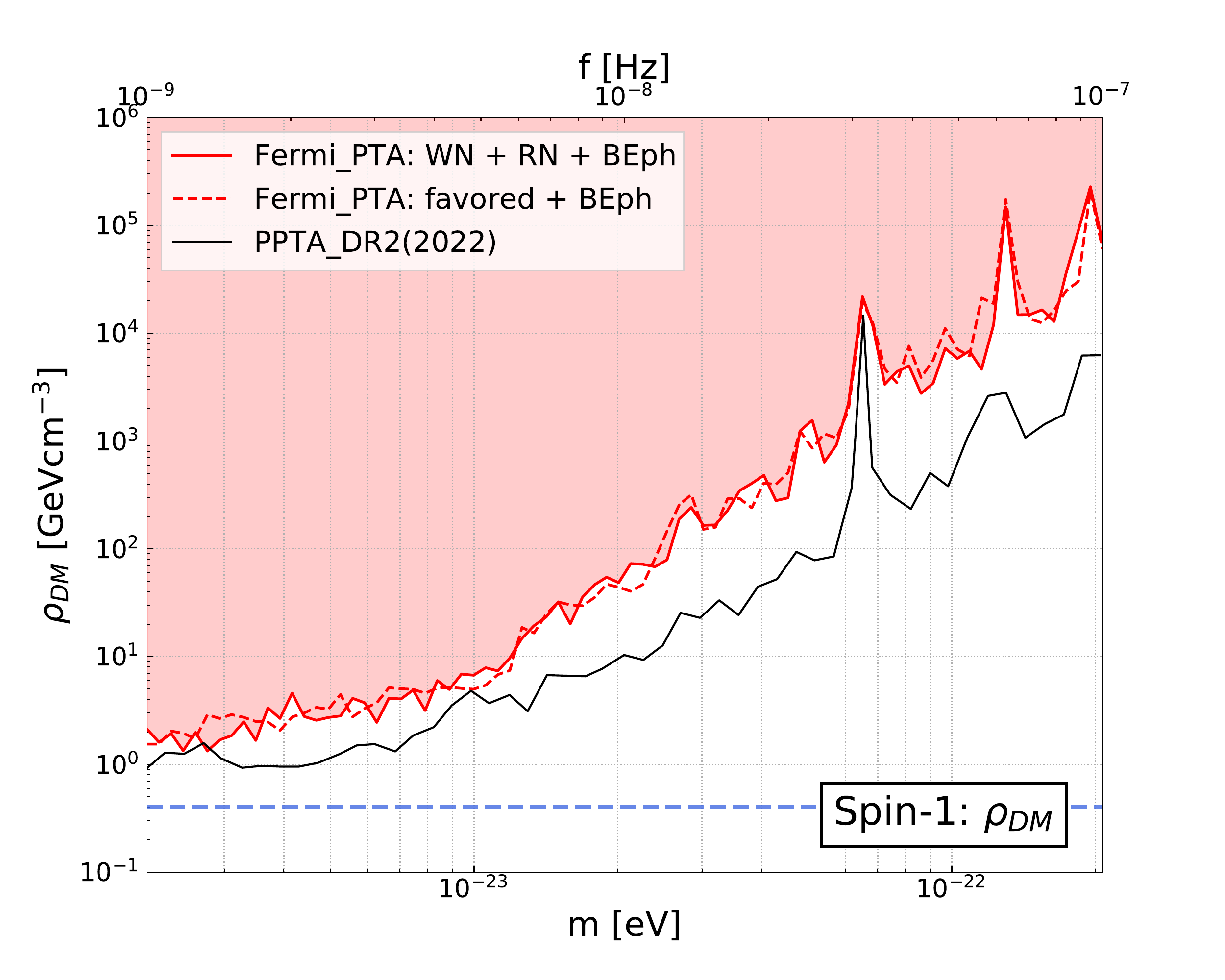}
\caption{Constraints on the local dark matter density $\rho_{\text{DM}}$ for the Spin-0/1 case with the gravitational effect: The left panel is for the Spin-0 ULDM and the black line represents upper limits from PPTA in 2018~\cite{1810.03227}. The right panel is for the Spin-1 ULDM and the previous upper limit based on PPTA DR2~\cite{2210.03880} is given as the black line.
For both panels, red solid lines are upper limits with all noise (WN $+$ RN $+$ BEph) components, while red dashed lines are for the case only including the favored noise model (as listed in Tab.~S2 in the Appendix Sec.B) and the BayesEphem uncertainty (favored $+$ BEph).
The blue dashed lines are on behalf of the case that the local dark matter density $\rho_{\text{DM}}$ equals to the current measurement $\rho_{0}=0.4$~GeV~cm$^{-3}$.} 
\label{Fig:upli_rho} 
\end{figure*}

\section{The Fermi-LAT PTA}\label{sec:FermiPTA}

\subsection{Data}\label{data}
The Fermi-LAT Collaboration uses 12.5 yr of Fermi-LAT data from 2008 Aug 04 to 2021 Jan 28 to form the first gamma-ray PTA, or called the Fermi-LAT PTA~\cite{2204.05226}.
It is composed of the 35 brightest and most stable gamma-ray MSPs.
However, only 29 of 35 MSPs in the Fermi-LAT PTA can efficiently estimate TOAs of pulses\footnote{ Another 6 MSPs (PSR J0312$-$0921, PSR J0418$+$6635, PSR J1513$-$2550, PSR J1543$-$5149, PSR J1741$+$1351, and PSR J1908$+$2105) do not reach the required log-likelihood threshold in the TOA estimation and can only be analyzed with the photon-by-photon method. Note that none of these 6 MSPs contribute significantly to the sensitivity of the Fermi-LAT PTA. See Supplementary Materials of Ref.~\cite{2204.05226} for more detail.}, and are suitable for the TOA-based method, which is widely used in radio PTAs.

In this work, we adopt the TOA-based method and use TOAs data recorded in the {\tt .tim} files for these 29 MSPs (as the {\tt Full 29} set in Ref.~\cite{2204.05226}).
We summarize the basic information~\cite{2204.05226, 2005ATNF} for each pulsar in Tab.~S2 in the Appendix Sec.B.

The Fermi-LAT PTA has calculated TOAs by a cadence of 2, 1.5, and 1 yr$^{-1}$ and obtained a total of 25, 19, and 12 TOAs for each pulsar.
The TOAs with a higher cadence can provide more high-frequency information and larger data volume.
Hence, in a general way, the 2 yr$^{-1}$ cadence is a relatively optimal choice.
However, PSR J0533+6759, PSR J0740+6620, PSR J1625-0021, PSR J1939+2134, PSR J2034+3632, and PSR J2256-1024 are relatively faint in all 29 MSPs and exhibit larger systematic errors for the 2 yr$^{-1}$ cadence than that for the longer integration~\cite{2204.05226}. 
For these 6 MSPs, reliable TOAs produced with the 1.5 yr$^{-1}$ cadence are adopted in the following analysis, the same as the preferred cadence used in Ref.~\cite{2204.05226}.

\subsection{Timing Model}\label{timingmodel}
TOAs of a pulsar can be reconstructed from three parts:
\begin{equation}\label{eq:tm}
{\rm TOA} \sim t_{\rm TM} + {\Delta t}_{\rm Noise} + s(t).
\end{equation}
The first term $t_{\rm TM}$ characterizes the timing model accounting for the pulsar ephemeris.
In this work, we use the optimized ephemerides model given by Ref.~\cite{2204.05226}\footnote{The corresponding {\tt .par} files can be found in \url{https://zenodo.org/record/6374291\#.YzVcbC-KFpR}.} and marginalize over model uncertainties.
The second term ${\Delta t}_{\rm Noise}$ is the timing residual contributed by noises.
The third term $s(t)$ is the timing residual induced by the signal for each ULDM effect described in Sec.~\ref{sec:ULDMmodel}.
The noise parameters, signal parameters, and their prior distributions are indicated in Tab.~S1 in the Appendix Sec.A.

The noises of PTA can be classified into three categories: the white (time-uncorrelated) noise (WN), the red (time-correlated) noise (RN), and the BayesEphem noise (BEph)~\cite{1801.02617}, respectively:
\begin{equation}\label{eq:noise}
{\Delta t}_{\rm Noise} = {\Delta t}_{\rm WN} + {\Delta t}_{\rm RN} + {\Delta t}_{\rm BEph}.
\end{equation}
We describe in detail these three noise components in the Appendix Sec.C.
Ref.~\cite{2204.05226} has made a single-pulsar noise analysis for all 29 pulsars and the favored noise model for each pulsar is listed in Tab.~S2 in the Appendix Sec.B.
For the sake of conservation, here we consider white and red noises for each pulsar and the BayesEphem noise for the PTA, and marginalize over their parameters in the likelihood analysis to search for the potential ULDM signal.

\section{Analysis and result}\label{sec:analysisresult}

\begin{figure*}
\centering 
\includegraphics[width=0.495\textwidth]{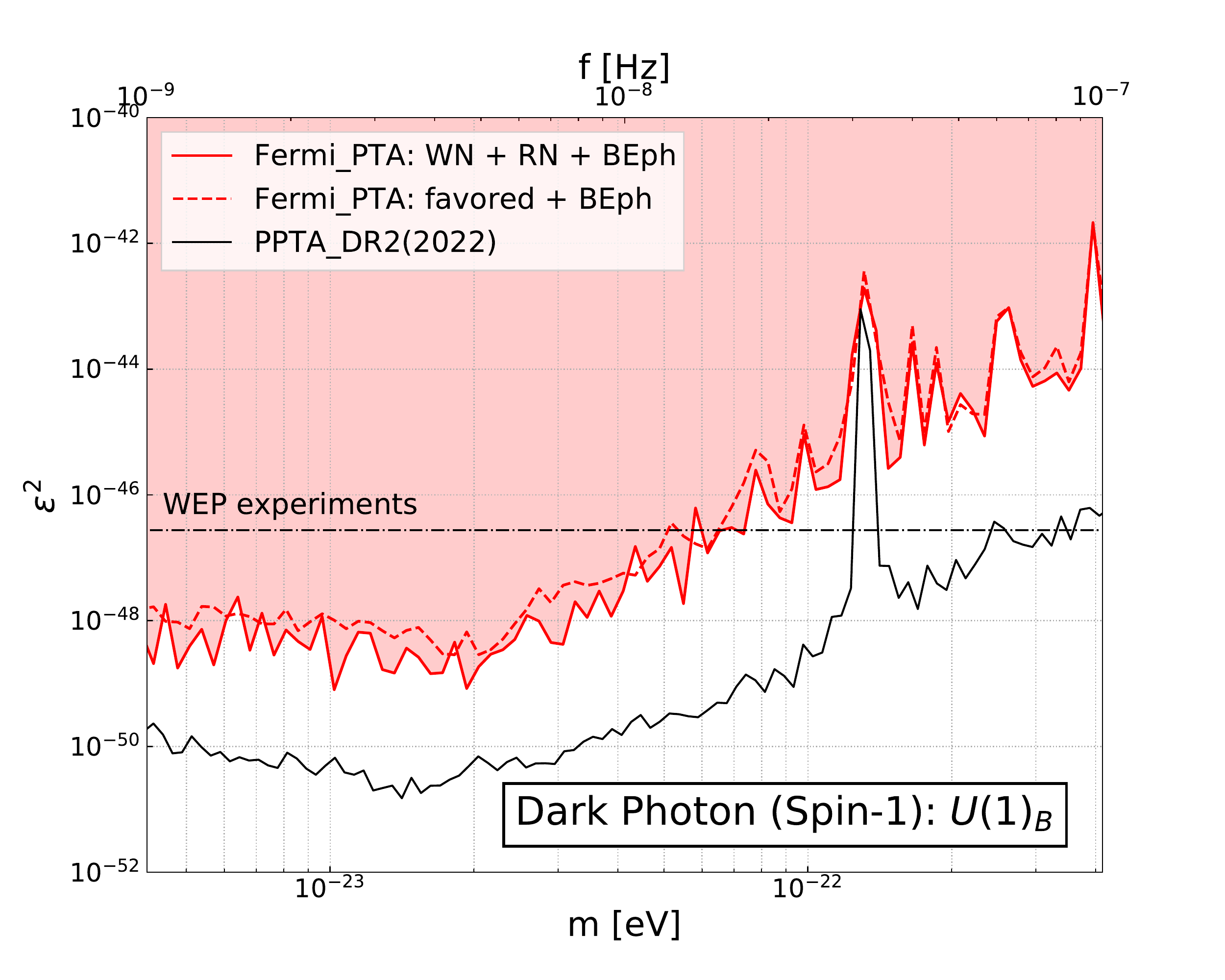}
\includegraphics[width=0.495\textwidth]{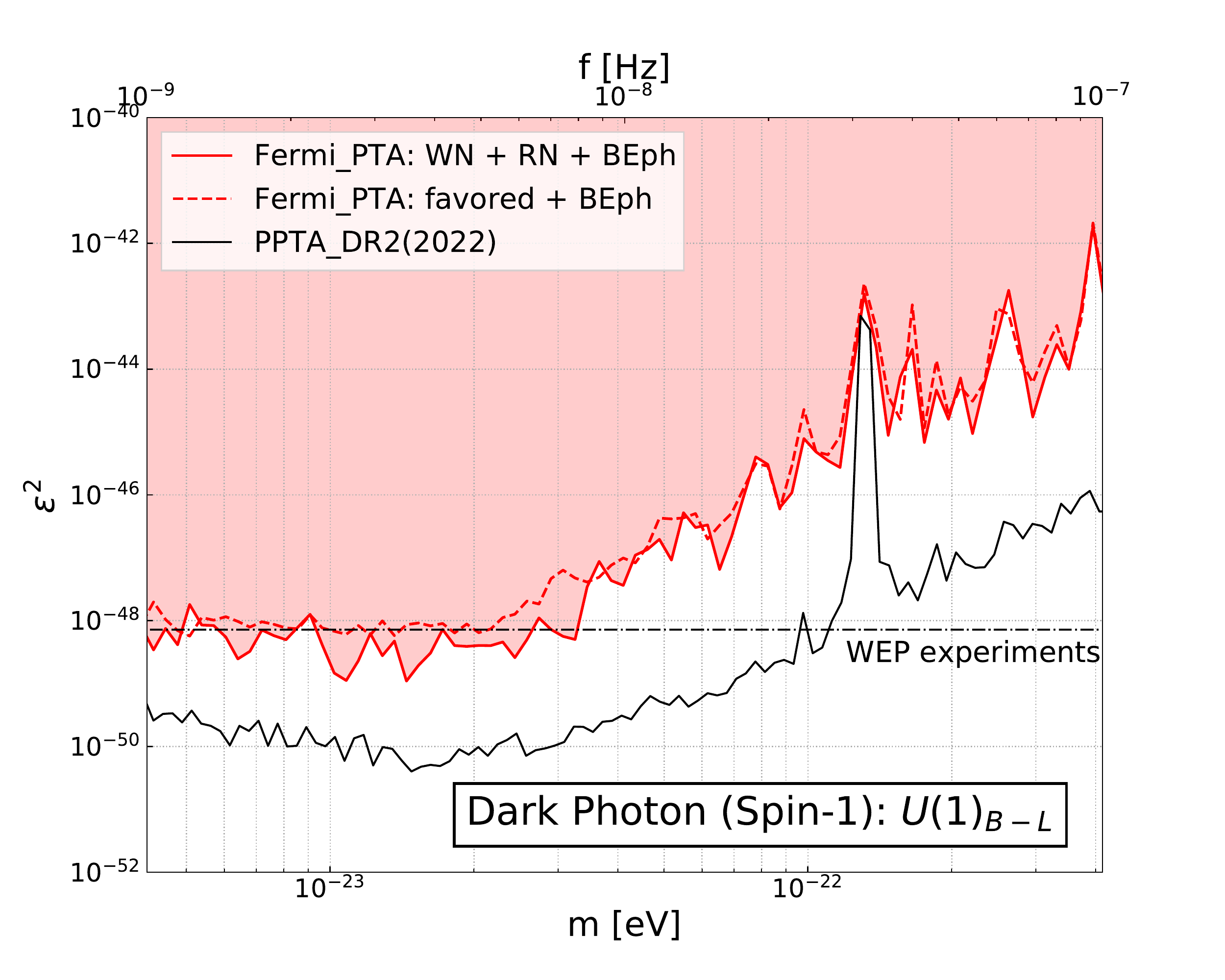}
\caption{Constraints on the fifth-force effect of the dark photon (Spin-1) model: The left panel shows the 95\% confidence level upper limits of the coupling strength $\epsilon$ for the $U(1)_{B}$ scenario, while the right panel is for the $U(1)_{B-L}$ scenario. 
The black lines show limits based on the PPTA DR2 dataset given in Ref.~\cite{2112.07687}.
The horizontal black dash-dotted lines are upper limits for the MICROSCOPE weak equivalence principle (WEP) experiment~\cite{1712.00483}. } 
\label{Fig:upli_spin1_DP} 
\end{figure*}

\subsection{Data analysis}\label{sec:data analysis}
To search for the ULDM signal with the Fermi-LAT PTA, we perform the likelihood analysis for two models: the signal model (labeled as ${\rm H_1}$) given in Eq.~(\ref{eq:tm}) and the null model (${\rm H_0}$) without timing residual contributed from the ULDM signal ($s(t)$). The {\tt PINT} software package\footnote{\url{https://github.com/nanograv/pint}}~\cite{PINT} and {\tt enterprise} package\footnote{\url{https://github.com/nanograv/enterprise}}~\cite{enterprise} are used to build the timing model of TOAs and perform the likelihood analysis. 
Then we use the {\tt PTMCMC} package\footnote{\url{https://github.com/jellis18/PTMCMCSampler}}~\cite{PTMCMC} as the stochastic sampling for posterior probabilities of parameters to obtain the best-fit values.
As for the ${\rm H_1}$ model, we repeat the analysis for each ULDM effect introduced in Sec.~\ref{sec:ULDMmodel}, respectively.
To compare models with and without signal, we take the likelihood ratio test between the ${\rm H_1}$ and ${\rm H_0}$ models.
The likelihood ratio (${\lambda_{\rm LR}}$) can be given by~\cite{Wilks1938}:
\begin{equation}
{\lambda_{\rm LR}}= 2 \ln \frac{\mathcal{L}_{\rm max}({\rm H_1})}{\mathcal{L}_{\rm max}({\rm H_0})},
\end{equation}
where $\mathcal{L}_{\rm max}({\rm H_1})$ and $\mathcal{L}_{\rm max}({\rm H_0})$ represent maximum likelihood values we get for ${\rm H_1}$ and ${\rm H_0}$, respectively.
According to Wilks’ theorem, the likelihood ratio~\cite{Wilks1938} follows the $\chi^{2}$ distribution with degrees of freedom equal to the difference in numbers of parameters for the ${\rm H_1}$ and ${\rm H_0}$ models.

Then we further derive upper limits on ULDM parameters for each effect, respectively. 
For a set of fixed ULDM mass $m$ in the corresponding frequency $f$ range ($10^{-9}$ Hz, $10^{-7}$ Hz), we sample posterior probabilities of ULDM parameters with the numerical marginalization for all noise (WN $+$ RN $+$ BEph) parameters mentioned above. 
When posterior probabilities in the range of [0, $x_{ul}$] equal to 95\%, we take $x_{ul}$ as the 95\% confidence level upper limit of one ULDM parameter ($x$) for the fixed ULDM mass $m$.
For completeness, we also calculated upper limits for the case only with favored noise models as listed in Tab.~S2 in the Appendix Sec.B. and the BayesEphem uncertainty, labeled as (favored $+$ BEph).

\subsection{Result}\label{sec:Result}

$\bullet$ {Spin-0: Gravitational Signal} --- Considering the gravitational effect of the Spin-0 ULDM, we get the likelihood ratio ${\lambda_{\rm LR, Spin-0}}$ equals to 3.9 with the best-fit parameters ($m$, $\Psi$) = ($1.38 \times 10^{-24}$ eV, $1.16 \times 10^{-19}$).
Considering the $\chi^2$ distribution with 32 degrees of freedom, the corresponding statistical significance is approximately to be zero.
Hence, the Fermi-LAT PTA shows no evidence of the gravitational signal from the Spin-0 ULDM, which is consistent with the result of PPTA~\cite{1810.03227}.
We calculate the 95\% confidence level upper limits of oscillation amplitude $\Psi$ for a set of fixed mass $m$ corresponding to the frequency from $10^{-9}$ Hz to  $10^{-7}$ Hz, which are shown in the left panel of Fig.~S1 in the Appendix Sec.D.
According to the Eq.~(\ref{eq:Psi-spin0}), we then conduct the 95\% upper limits of local dark matter density $\rho_{\rm DM}$ and exhibit them in the left panel of Fig.~\ref{Fig:upli_rho}.
Our constraints on the local dark matter density $\rho_{\rm DM}$ is larger than the observed local dark matter density $\rho_{0}=0.4$~GeV~cm$^{-3}$ and higher than upper limits obtained by PPTA in 2018~\cite{1810.03227}.

$\bullet$ {Spin-1: Gravitational Signal} ---
As for the gravitational signal of the Spin-1 case, 
the best-fit parameters are ($m$, $h_{\rm osc}$) = ($1.09 \times 10^{-23}$ eV, $6.11 \times 10^{-16}$), and the corresponding likelihood ratio is ${\lambda_{\rm LR,Spin-1}} = 3.8$ following the $\chi^2$ distribution with 34 degrees of freedom.
Same as above, we find no significant evidence for the Spin-1 gravitational signal, which is consistent with result from PPTA DR2~\cite{2210.03880}.
The 95\% confidence level upper limits with the Fermi-LAT PTA for oscillation amplitude $h_{\rm osc}$ is displayed in the right panel of Fig.~S1 in the Appendix Sec.D.
Conducted from the Eq.~(\ref{eq:h-spin1}), upper limit of $\rho_{\rm DM}$ is shown in the right panel of Fig.~\ref{Fig:upli_rho}.

$\bullet$ {Spin-1: Fifth-Force Signal} --- 
In the case of dark photon, we consider two new interaction ($U(1)_{B}$ and $U(1)_{B-L}$) scenarios.
For $U(1)_{B}$ / $U(1)_{B-L}$ scenario, best-fit parameters we get are (m, $\epsilon$) $=$ ($6.76 \times 10^{-22}$ eV, $9.95 \times 10^{-26}$) / ($1.36 \times 10^{-22}$ eV, $5.03 \times 10^{-23}$), respectively.
Corresponding likelihood ratios of ${\lambda_{\rm LR,DP}} = 2.1$ / $25.2$ are asymptotically the $\chi^2$ distribution with 37 degrees of freedom and have a statistical significance of $\sim $ 0.
The 95\% confidence level upper limits of coupling strength $\epsilon$ for the $U(1)_{B}$ and $U(1)_{B-L}$ scenarios are presented in the left and right panels of Fig.~\ref{Fig:upli_spin1_DP}, respectively.
Our constraints are found weaker than upper limits from PPTA DR2~\cite{2112.07687}.
Especially for the $U(1)_{B}$ scenario with masses below $\sim 7 \times  10^{-23}$ eV, we set stronger constraints than limits from the MICROSCOPE WEP experiment~\cite{1712.00483} (the horizontal black dash-dot line in Fig.~\ref{Fig:upli_spin1_DP}).

\begin{figure}
\centering 
\includegraphics[width=0.495\textwidth]{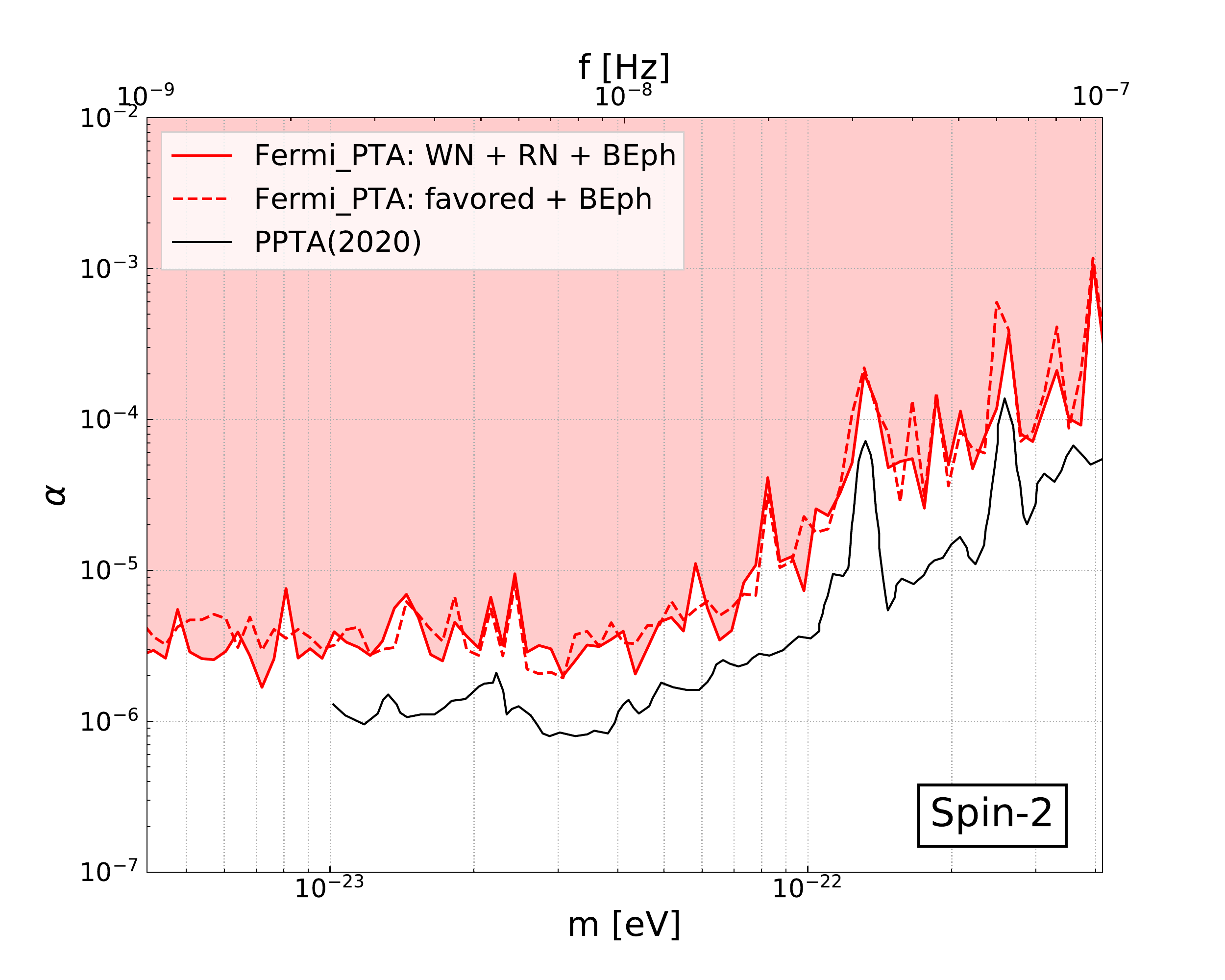}
\caption{The 95\% confidence level upper limits of the coupling strength $\alpha$ for the Spin-2 ULDM model. 
The black line indicates the upper limit set by the PPTA data in Ref.~\cite{2005.03731}.} 
\label{Fig:upli_spin2} 
\end{figure}

$\bullet$ {Spin-2: Modified Gravitational Signal} --- 
As for the Spin-2 ULDM model, we obtain the best-fit parameters as (m, $\alpha$) $=$ ($1.43 \times 10^{-23}$ eV, $1.45 \times 10^{-6}$) with the likelihood ratio of ${\lambda_{\rm LR, Spin-2}} = 7.3$.
Following the $\chi^2$ distribution with 32 degrees of freedom, this likelihood ratio corresponds to a significance close to 0. 
The 95\% confidence level upper limits of coupling strength $\alpha$ are further derived for a set of fixed masses of the Spin-2 ULDM as shown in Fig.~\ref{Fig:upli_spin2}, which is also weaker than that from PPTA ~\cite{2005.03731}.

\section{Discussion and Summary}\label{sec:Summary}

ULDM is a theoretical form of dark matter that consists of extremely light particles with masses of $\sim 10^{-22}$ eV beyond the standard model. 
PTA is a valuable tool for searching for these ULDMs.
The idea behind this is that the (gravitational or fifth-force) interactions between ULDM particles and standard model particles could cause slight variations in the pulse arrival times, which can be precisely measured by observing the pulsars over long periods of time. 

In this letter, we first use the gamma-ray PTA constructed from the Fermi-LAT observation of 29 MSPs to indirectly detect ULDM.
Compared with the radio PTAs, the gamma-ray PTA is not affected by the turbulent IISM along the line of sight which could induce remarkable red noises for the radio PTAs.
Utilizing the traditional TOA-based method, here we search for four ULDM models including the gravitational effects for ULDM particles with the different Spin-0/1, the coupling (fifth-force) effect of dark photon, and the modified gravitational effect for the Spin-2 ULDM.
For all four kinds of ULDM effects, we don't find any significant signal based on the Fermi-LAT PTA dataset, which agrees with the non-detection of ULDM signal from radio PTAs and direct detection experiment.
Then we further set limits on parameters of ULDM with the 95\% confidence level in the corresponding frequency $f$ range of ($10^{-9}$ Hz, $10^{-7}$ Hz).
As for the Spin-0/1 gravitational effects, our constraints are larger than the measured local dark matter density $\rho_{0}=0.4$~GeV~cm$^{-3}$ and thus hard to further limit the particle mass of ULDM.
In the case of the Spin-1 dark photon effect, our constraints for the $U(1)_{B}$ scenario with masses below $\sim 7 \times  10^{-23}$ eV is more competitive than that of the MICROSCOPE WEP experiment. 
For all four effects, our limits based on the gamma-ray PTA are weaker than those of radio PTAs, due to the relatively smaller amounts of timing data.  
However, our work could be an independent and complementary check for the ULDM detection based on radio PTAs.

In the future, the Very Large Gamma-ray Space Telescope (VLAST)~\cite{2022AcASn..63...27F}, one of the next generation space-based gamma-ray telescopes with an effective detection area of $\sim 4~{\rm m^{2}}$, will be useful to establish a stronger gamma-ray PTA to probe the ULDM and the nanohertz-frequency gravitational waves. 
As a very rough estimate, assuming that VLAST can detect the same number of pulsars with the same TOA precision, but enlarge the number of TOAs by a factor of $5$, we find that the upper limit of the oscillation amplitude $\Psi$ will be smaller by a factor of $\sim\sqrt{5}$. Additional improvement may be achieved if the timing model and noise model can be improved given larger photon statistics, which can be explored in detail in the future.

\begin{acknowledgments}
This work is supported by the National Key Research and Development Program of China (No. 2022YFF0503304), the Natural Science Foundation of China (No. 12003069, 11921003), the Chinese Academy of Sciences, and the Entrepreneurship and Innovation Program of Jiangsu Province.
\end{acknowledgments}

\section*{Appendix}

\setcounter{figure}{0}
\renewcommand\thefigure{S\arabic{figure}}
\setcounter{table}{0}
\renewcommand\thetable{S\arabic{table}}

\begin{table*}
\centering
\caption{The parameters of signal and noise models for the analysis.}
\label{tab:parameter}
\begin{tabular}{c|ccc}
\hline\hline
Model & Parameter & Prior & Description \\
\hline
{\shortstack{Spin-0: Gravitational Signal}}
&$\theta_p$ & U[$0$,$2\pi$] &  Phase for the pulsar term, one per pulsar\\
&$\theta_e$  & U[$0$,$2\pi$] &  Phase for the Earth term, one per PTA\\
&$\Psi$  & log-U[$10^{-20}$,$10^{-12}$] & Oscillation amplitude for the Spin-0 ULDM field, one per PTA\\
&$m$ & log-U[$10^{-24}$,$10^{-21}$] eV & Mass for the Spin-0 ULDM, one per PTA\\
\hline
{\shortstack{Spin-1: Gravitational Signal}}
&$\theta_p$ & U[$0$,$2\pi$] &  Phase for the pulsar term, one per pulsar\\
&$\theta_e$  & U[$0$,$2\pi$] &  Phase for the Earth term, one per PTA\\
&$\theta_{\rm osc}$  & U[$0$,$\pi$] & Polar angle of the Spin-1 ULDM oscillation direction, one per PTA\\
&$\phi_{\rm osc}$  & U[$0$,$2\pi$] & Azimuth angle of the Spin-1 ULDM oscillation direction, one per PTA\\
&$h_{\rm osc}$  & log-U[$10^{-19}$,$10^{-9}$] & Oscillation amplitude for the Spin-1 ULDM field, one per PTA\\
&$m$ & log-U[$10^{-24}$,$10^{-21}$] eV & Mass for the Spin-0 ULDM, one per PTA\\
\hline
{\shortstack{Spin-1: Fifth-Force Signal}}
&$\theta_p$ & U[$0$,$2\pi$] & Phase for the pulsar term, one per pulsar\\
&$\theta_e^{i}$  & U[$0$,$2\pi$] & Phase for the Earth term, three per PTA\\
&$(\tilde{A}_0^{i})^2$ &$f(x)=e^{-x}$& 
The square of normalized dark photon amplitude, three per PTA\\\
&$\epsilon$  & log-U[$10^{-28}$,$10^{-16}$] & Coupling strength for the dark photon field, one per PTA\\
&$m$ & log-U[$10^{-24}$,$10^{-21}$] eV & Mass for the dark photon, one per PTA\\
\hline
{\shortstack{Spin-2: Modified Gravitational Signal}}
&$\theta_p$ & U[$0$,$2\pi$] & Phase for the pulsar term, one per pulsar\\
&$\theta_e$  & U[$0$,$2\pi$] & Phase for the Earth term, one per PTA\\
&$\alpha$  & log-U[$10^{-12}$,$10^{-1}$] & Coupling strength for the Spin-2 ULDM field,  one per PTA\\
&$m$ & log-U[$10^{-24}$,$10^{-21}$] eV & Mass for the Spin-2 ULDM, one per PTA\\
\hline
{\shortstack{White Noise (WN)}}
&{\tt EFAC}  & U[$0.1$,$5$] & Re-scaling factor, one per pulsar\\
&{\tt EQUAD} & log-U [$10^{-9}$,$10^{-5}$] & Extra white noise, one per pulsar\\
\hline
{\shortstack{Red Noise (RN)}}
&$A$ & log-U[$10^{-18}$,$10^{-9}$] & Amplitude for red noise, one per pulsar\\
&$\gamma$ & U[0,7] & Index for red noise, one per pulsar\\
\hline
{\shortstack{BayesEphem Noise (BEph)}}
&$z_{\rm drift}$ &U[$-10^{-9}$,$10^{-9}$] rad yr$^{-1}$ & Drift-rate of Earth’s orbit about ecliptic z-axis,one per PTA \\
&$\Delta M_{\rm Jupiter}$&N$(0,1.5\times10^{-11})$ $M_{\odot}$& Perturbation of Jupiter’s mass, one per PTA\\
&$\Delta M_{\rm Saturn}$&N$(0,8.2\times10^{-12})$ $M_{\odot}$& Perturbation of Saturn’s mass, one per PTA\\
&$\Delta M_{\rm Uranus}$&N$(0,5.7\times10^{-11})$ $M_{\odot}$& Perturbation of Uranus’ mass, one per PTA\\
&$\Delta M_{\rm Neptune}$&N$(0,7.9\times10^{-11})$ $M_{\odot}$& Perturbation of Neptune’s mass, one per PTA\\
&${\rm PCA}_i$&U[$-0.05$,$0.05$]& Principal components of Jupiter’s orbit, six per PTA\\
\hline\hline
\end{tabular}
\end{table*}

\subsection{Ultralight dark matter model}\label{sec:UDMmodel}

\subsubsection{Spin-0: Gravitational Signal}
With a macroscopic de Broglie wavelength of galactic scales, the Spin-0 ULDM particles around the Earth-pulsars system would produce the periodic oscillation in gravitational potentials, which could cause sinusoidal variations in the times of arrival (TOA) of pulse~\cite{1309.5888}.
This oscillating effect is similar to the steady monochromatic gravitational wave from a distant source.
The Spin-0 ULDM could induce a time-dependent gravitational potential with the frequency ($f$) twice the mass of ULDM particle ($m$)~\cite{1309.5888}: 
\begin{equation}
f = \frac{2m}{2\pi} = 4.8 \times 10^{-9} {\rm Hz} \left(\frac{m}{10^{-23}{\rm eV}}\right)\,.
\label{eq:f-spin0}
\end{equation}
The corresponding timing residuals of TOAs can be written as~\cite{1810.03227}: 
\begin{eqnarray}
s(t) &=& \frac{\Psi_e}{2\pi f}\sin(2\pi ft+2\theta_{e})-\frac{\Psi_p}{2\pi f}\sin\left[2\pi f \left(t-\frac{d_{p}}{c}\right)+2\theta_p^{'}\right] \nonumber \\
&=& \frac{\Psi_e}{2\pi f}\sin(2\pi ft+2\theta_{e})-\frac{\Psi_p}{2\pi f}\sin\left[2\pi f t+2\theta_p\right],
\label{eq:st0-spin0}
\end{eqnarray}
where $d_{p}$ is the distance from the pulsar in the Fermi-LAT PTA to the Earth, $\Psi_{e}$ and $\Psi_{p}$ are the amplitude of the oscillating part of the gravitational potential for the Earth and pulsar terms, recpectively. $\theta_{e}$ and $\theta_{p}^{'}$ each represent the scalar-field oscillation phases for the Earth and pulsar terms. 
To simplify the formula, we take the phase of pulsar term as $\theta_{p}$ which absorbs the time difference ($d_{p}/c$) between a pulsar and the Earth.

The amplitude is related to the density of dark matter $\rho_{\text{DM}}$,
\begin{equation}
\Psi=\frac{G \rho_{\text{DM}}}{\pi f^2}\approx 6.1 \times 10^{-18} \left(\frac{m}{10^{-22}\,\text{eV}}\right)^{-2} \left(\frac{\rho_{\text{DM}}}{\rho_{0}}\right)\, ,
\label{eq:Psi-spin0}
\end{equation}
where $\rho_{0}=0.4 \, \text{GeV cm}^{-3}$ is the measured local dark matter density near the Earth~\cite{1205.4033,1708.07836}.
Since the pulsar distance $d_{p}$ to the Earth are relatively small (for 27 of 29 pulsars below 3 kpc), the dark matter density around pulsars can be approximately considered to be the local density near the Earth.
Hence the amplitude for the Earth term and pulsar terms are set to be the same (as implicitly done in Ref~\cite{1810.03227}), uniformly labeled as $\Psi$.
Under this assumption and the Eq.~(\ref{eq:f-spin0}), the timing residuals of TOAs can be written into
\begin{equation}
s(t)=\frac{\Psi}{m}\sin(\theta_{e}-\theta_{p})\cos(2 m t+\theta_{e}+\theta_{p})\, .
\label{eq:st1-spin0}
\end{equation}
The prior distributions of $\theta_{p}$ and $\theta_{p}$ are set to be uniform between 0 to 2$\pi$ as listed in Tab.~\ref{tab:parameter}.
The whole derivation of Eq.~(\ref{eq:st1-spin0}) is presented in Refs.~\cite{1309.5888, 1810.03227}.

\subsubsection{Spin-1: Gravitational Signal}
Same with the Spin-0 model,
the oscillation of gravitational potential induced by the Spin-1 ULDM could also result in periodic timing signals in observations of PTAs, with the frequency twice the ULDM mass as in the Eq.~(\ref{eq:f-spin0}).
However, the Spin-1 gravitational timing signals are also related to the oscillation direction of the vector field, which is a distinctive feature different from the Spin-0 case. 
This kind of signal has been derived in detail in the Refs~\cite{1912.10210,2210.03880}, which can be generally divided into two parts. One is $s_{\Psi}(t)$ for the oscillation in gravitational potentials:
\begin{equation}
s_{\Psi}(t)=\frac{\Psi_{\mathrm{osc}}}{m}\sin(\theta_{e}-\theta_{p})\cos(2 m t+\theta_{e}+\theta_{p})\,,
\label{eq:st0-spin1}
\end{equation}
where $m$ is the mass of the Spin-1 ULDM particle, $\Psi_{\mathrm{osc}}$ represents the corresponding amplitude of potential, $\theta_{e}$ and $\theta_{p}$ are oscillation phases for the Earth and pulsar terms, respectively. 
The other is $s_{h}(t)$ for the traceless spatial metric perturbations, which exhibits the anisotropy for additional degrees of freedom in the vector field and does not exist in the Spin-0 case:
\begin{equation}
s_{h}(t)=-\frac{1}{4}\left(1+3\cos2\theta\right)\frac{h_{\mathrm{osc}}}{m}\sin(\theta_{e}-\theta_{p})\cos(2 m t+\theta_{e}+\theta_{p})\,, 
\label{eq:st1-spin1}
\end{equation}
where $h_{\mathrm{osc}}$ is the amplitude of perturbations, $\theta$ is the angle between the line of sight to a pulsar and the oscillation direction of the vector ULDM ($\theta_{\rm osc}$, $\phi_{\rm osc}$).
The amplitudes for these two oscillation parts also depend on the local dark matter density $\rho_{\text{DM}}$ and can be given as, respectively:
\begin{equation}
\Psi_{\mathrm{osc}}=-\frac{\pi G \rho_{\text{DM}}}{3 m^2}\approx -2.2 \times 10^{-18} \left(\frac{m}{10^{-22}\,\text{eV}}\right)^{-2} \left(\frac{\rho_{\text{DM}}}{\rho_{0}}\right)\,,
\label{eq:Psi-spin2}
\end{equation}

\begin{equation}
h_{\mathrm{osc}}=\frac{8 \pi G \rho_{\text{DM}}}{3 m^2}\approx 1.7 \times 10^{-17} \left(\frac{m}{10^{-22}\,\text{eV}}\right)^{-2} \left(\frac{\rho_{\text{DM}}}{\rho_{0}}\right)\,,
\label{eq:h-spin1}
\end{equation}
where the measured local dark matter density is taken as $\rho_{0}=0.4 \, \text{GeV cm}^{-3}$ (same as above).
Combining these two parts of oscillation, we get the total Spin-1 gravitational timing signal written as
\begin{eqnarray}
s(t) &=& s_{\Psi}(t) + s_{h}(t) \nonumber \\
&=& -\frac{3}{8}\left(1+2\cos2\theta\right)\frac{h_{\mathrm{osc}}}{m}\sin(\theta_{e}-\theta_{p})\cos(2 m t+\theta_{e}+\theta_{p})\,,
\nonumber \\
\label{eq:st2-spin1}
\end{eqnarray}
which is angle-dependent, unlike the Spin-0 model.
When the vector field oscillates along with the line-of-sight direction, the maximal amplitude was found as three times larger than that of the Spin-0 case~\cite{1912.10210}.
With the PPTA second data release (DR2) dataset, the Ref.~\cite{2210.03880} has not found a preferred oscillation direction of the Spin-1 ULDM ($\theta_{\rm osc}$, $\phi_{\rm osc}$), thus we also set them as free parameters in our work.

\subsubsection{Spin-1: Fifth-Force Signal}
As a special case of the Spin-1 ULDM candidates, the dark photon is the hypothetical gauge boson of a $U(1)$ interaction proposed by the string theory~\cite{0805.4037,0909.0515,1103.3705}.
The mass of dark photon could be generated by the Higgs mechanism or the Stueckelberg mechanism, and it is naturally light. 
If the $U(1)$ symmetry is for the baryon number (the $U(1)_B$ interaction) or for the baryon-minus-lepton number (the $U(1)_{B-L}$ interaction), there would be the ``fifth-force" interaction with the ordinary matter, which could be used to probe the aark photon. Ref.~\cite{2112.07687} has compared the pure gravitational effect for the vector ULDM with the ``fifth-force" effect and find the ``fifth force" could produce larger timing residuals when $\epsilon\times m/{\rm eV}>10^{-48.35}$ for the $U(1)_B$ scenario and $10^{-48.24}$ for the $U(1)_{B-L}$ scenario, where $\epsilon$ is the coupling strength and $m$ is the dark photon mass. 
Here we consider the ``fifth-force" effect for the dark photon case, instead of the previously introduced gravitational effect.

The frequency of the oscillating signal in timing residuals caused by the ``fifth force" is given by $m$:
\begin{equation}
f = \frac{m}{2\pi} = 2.4 \times 10^{-9} {\rm Hz} \left(\frac{m}{10^{-23}{\rm eV}}\right)\,.
\label{eq:f-spin1}
\end{equation}
And the ``fifth force" induced by dark photon with the mass of $m$ can lead to an acceleration on the test mass $m_{\rm t}$ located at $\boldsymbol{x}$~\cite{1801.10161,2112.07687}:
\begin{equation}
\boldsymbol{a}(t,\boldsymbol{x})\simeq \epsilon e 
\frac{q}{m_{\rm t}}m\boldsymbol{A_0}\cos{\big(mt-\boldsymbol{k \cdot x}+\boldsymbol{\theta}(\boldsymbol{x})\big)},
\label{eq:a-spin1}
\end{equation}
where the coupling strength of the new $U(1)$ gauge interaction $\epsilon$ is normalized to the electromagnetic coupling constant $e$, $q$ counts the number of the baryon ($B$) or the neutron (baryon-minus-lepton, $B-L$) and $\boldsymbol{k}$ represents the characteristic momentum.
$\boldsymbol{A_0}$ and $\boldsymbol{\theta}$ are on behalf of the gauge potential and phase of the dark photon field, respectively.
The amplitude of $\boldsymbol{A_0}$ depends on the dark matter density which we set to the observed local dark matter density $\rho_{0}=0.4$~GeV~cm$^{-3}$.
The averaged value of the amplitude can be given by $|\boldsymbol{A_0}|^2 = 2\rho_{0}/m^2$.
And we assume there is a common amplitude $\boldsymbol{A_0}$ of dark photon field around the Earth and pulsars.
The displacement caused by the acceleration can be approximately given as 
\begin{equation}
\Delta\boldsymbol{x}(t,\boldsymbol{x})= - \frac{\epsilon eq}{m_t m}
\boldsymbol{A_0}\cos{\big(mt-\boldsymbol{k \cdot x}+\boldsymbol{\theta}(\boldsymbol{x})\big)}.
\label{displacement}
\end{equation}

Hence, the total periodic timing residuals caused by the ``fifth-force" effect is a combination of the Earth (labeled as $_e$) term and pulsar (labeled as $_p$) term, which has been derived in our previous work \cite{2112.07687}:
\begin{eqnarray}
\label{eq:stb-spin1}
s(t)^{(B)} &=&  \frac{\epsilon e}{m} \boldsymbol{A_0} \left[ \frac{q^{(B)}_e}{m_e} \cos\left(m t+\boldsymbol{\theta}_e\right)\right. - \nonumber \\
& &\left.\frac{q^{(B)}_p}{m_p} \cos(m t+\boldsymbol{\theta}_p)  \right]\cdot \boldsymbol{n}, \\
\label{eq:stbl-spin1}
s(t)^{(B-L)} &=&  \frac{\epsilon e}{m} \boldsymbol{A_0} \left[ \frac{q^{(B-L)}_e}{m_e} \cos\left(m t+\boldsymbol{\theta}_e\right)\right. - \nonumber \\
& &\left.\frac{q^{(B-L)}_p}{m_p} \cos(m t+\boldsymbol{\theta}_p)  \right]\cdot \boldsymbol{n},
\end{eqnarray}
where $\boldsymbol{n}$ is the normalized position vector pointing from the Earth to the pulsar, $q^{(B)}_{e,p}$, $q^{(B-L)}_{e,p}$, and $m_{e,p}$ each represent the $B$ number, the $B-L$ number, the mass of the Earth and the pulsar. For the $U(1)_B$ interaction, $q^{(B)}/m$ can be roughly taken as $1/{\rm GeV}$ for both the Earth and the pulsar. While for the $U(1)_{B-L}$ scenario, $q^{(B-L)}/m$ is approximately $0.5/{\rm GeV}$ for the Earth and $1/{\rm GeV}$ for the pulsar.
The amplitude $\boldsymbol{A_0}$ consists of three spatial components labeled as ${A}_0^{i}$. 
For each spatial component, we define a normalized amplitude $\tilde{A}_0^{i}$ by
\begin{equation}
\tilde{A}_0^{i}= \sqrt{\frac{3m^2}{2\rho_{0}}} {A}_0^{i},
\end{equation}
which cancels the constant part $ \sqrt{2\rho_0/3m^2}$ of $A_0^i$, becoming a dimensionless amplitude.
Simulation shows that $(\tilde{A}_0^{i})^2$ follows an exponential distribution \cite{2112.07687}, which is used as prior, as given in Tab.~\ref{tab:parameter}.

\subsubsection{Spin-2: Modified Gravitational Signal}
The massive Spin-2 field is proposed as a modification of gravity and behaves as a potential dark matter candidate~\cite{Aoki:2017cnz,Marzola:2017lbt}.
A direct coupling with the energy-momentum tensor of the standard matter is naturally involved by the tensorial structure and action of the Spin-2 ULDM, which can be parametrized by the dimensionless strength constant $\alpha$.
The oscillating frequency $f$ for the Spin-2 case is given by $m$, which is the same as the Eq.~(\ref{eq:f-spin1}) for the fifth-force effect of dark photon.
Based on the bimetric gravity, Ref.~\cite{2005.03731} detailedly describes the modified gravitational effect of any universally-coupled Spin-2 ULDM on the PTA.
For the Spin-2 case, photons from the pulsar along the direction of propagation would follow the geodesics of the new metric relying on the Spin-2 ULDM field.
Here we approximately take $\rho_{0}=0.4 \, \text{GeV cm}^{-3}$ as dark matter density $\rho_{\rm DM}$ for both the pulsar term and the Earth term.
Conducted from the Eq.(A.10) in Ref.~\cite{2005.03731}, we obtained the time residual as:
\begin{eqnarray}
\label{eq:st-spin2}
s(t) &=& \frac{\alpha \sqrt{2\rho_{0}}}{\sqrt{15} m^2 {M_{\text{P}}}}\cos\left(mt+\frac{\theta_{e}+\theta_{p}}{2}\right)\, ,
\end{eqnarray}
where ${M_{\text{P}}}$ is the reduced Planck mass, and $\theta_{p}$ includes the time difference ($d_p/c$) from a pulsar to the Earth.
Prior distributions of Spin-2 model parameters are summarized in Tab.~\ref{tab:parameter}.

\begin{table}[!h]
\centering
\caption{The basic information for 29 millisecond pulsars in the Fermi-LAT PTA}
\vspace{0.5cm}
\begin{tabular}{c|c c c c c}
\hline
\hline
\multicolumn{1}{c|}{Pulsar Name} & \multicolumn{1}{c}{Cadence}  & \multicolumn{1}{c}{$d_p$} & \multicolumn{1}{c}{$\sigma_{\rm TOA}$} & \multicolumn{1}{c}{rms} & Noise Model\\
\multicolumn{1}{c|}{(PSR)}  & \multicolumn{1}{c}{(TOA/yr)} & \multicolumn{1}{c}{(kpc)} & \multicolumn{1}{c}{($\mu$s)} & \multicolumn{1}{c}{($\mu$s)} & \multicolumn{1}{c}{(favored)}\\
\hline
J0030$+$0451 & 2   & 0.32 & 4.13  & 4.32  & None\protect\footnote{``None'' means that there is no additional noise found and the white noise equals to measured TOA uncertainties.}  \\ 
J0034$-$0534 & 2   & 1.35 & 13.41 & 10.11 & None  \\ 
J0101$-$6422 & 2   & 1.00 & 14.16 & 16.85 & None  \\ 
J0102$+$4839 & 2   & 2.31 & 16.14 & 19.18 & None  \\
J0340$+$4130 & 2   & 1.60 & 18.26 & 23.03 & None  \\ 
J0533$+$6759 & 1.5 & 2.40 & 16.80 & 13.30 & None  \\ 
J0613$-$0200 & 2   & 0.78 & 19.41 & 14.46 & None  \\ 
J0614$-$3329 & 2   & 0.63 & 2.53  & 2.71  & None  \\ 
J0740$+$6620 & 1.5 & 1.15 & 10.53 & 2.34  & None  \\ 
J1124$-$3653 & 2   & 0.99 & 10.33 & 3.13  & None  \\ 
J1231$-$1411 & 2   & 0.42 & 2.43  & 2.51  & None  \\ 
J1514$-$4946 & 2   & 0.91 & 12.48 & 14.22 & None  \\
J1536$-$4948 & 2   & 0.98 & 7.02  & 7.01  & None  \\
J1614$-$2230 & 2   & 0.70 & 5.80  & 10.35 & None  \\
J1625$-$0021 & 1.5 & 0.95 & 18.52 & 15.27 & None  \\
J1630$+$3734 & 2   & 1.19 & 5.82. & 5.64  & None  \\
J1810$+$1744 & 2   & 2.36 & 10.26 & 12.44 & None  \\
J1816$+$4510 & 2   & 4.36 & 19.72 & 25.01 & None  \\
J1858$-$2216 & 2   & 0.92 & 13.10 & 19.05 & None  \\
J1902$-$5105 & 2   & 1.65 & 7.84  & 7.62  & None  \\
J1939$+$2134 & 1.5 & 3.50 & 7.20  & 11.45 & None  \\
J1959$+$2048 & 2   & 1.40 & 4.78  & 9.35  &   WN  \\
J2017$+$0603 & 2   & 1.40 & 14.38 & 19.94 & None  \\
J2034$+$3632 & 1.5 & $-$   & 17.59 & 18.73 & None  \\
J2043$+$1711 & 2   & 1.39 & 8.04  & 9.29  & None  \\
J2214$+$3000 & 2   & 0.60 & 20.09 & 23.94 & None  \\
J2241$-$5236 & 2   & 1.04 & 5.30  & 6.57  & WN    \\
J2256$-$1024 & 1.5 & 2.08 & 4.13  & 5.23  & None  \\
J2302$+$4442 & 2   & 0.86 & 11.38 & 8.20  & None  \\
\hline 
\hline
\end{tabular}
\label{tab:29psr}
\end{table}

\begin{figure*}[!htb]
\centering
\includegraphics[width=0.495\textwidth]{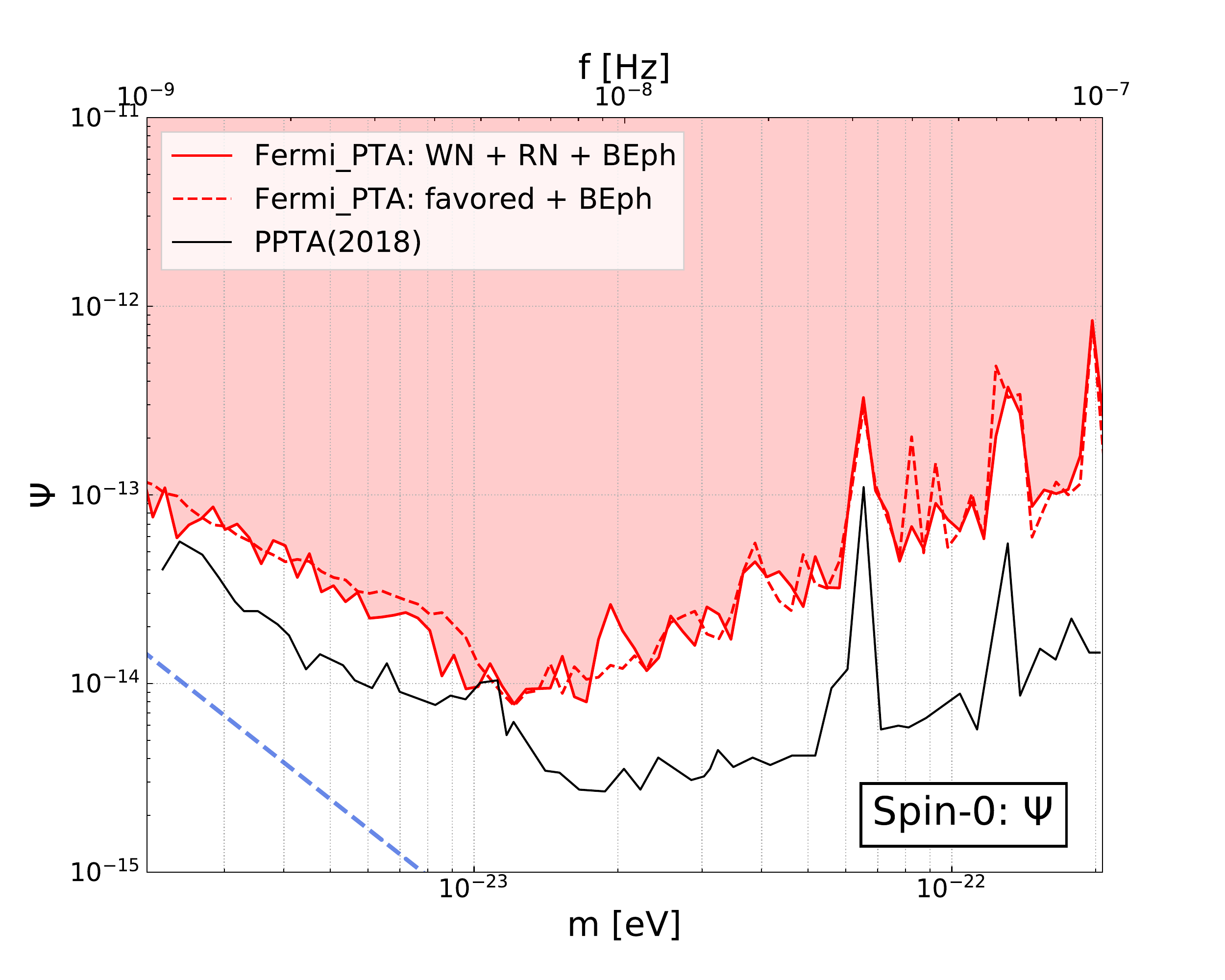}
\includegraphics[width=0.495\textwidth]{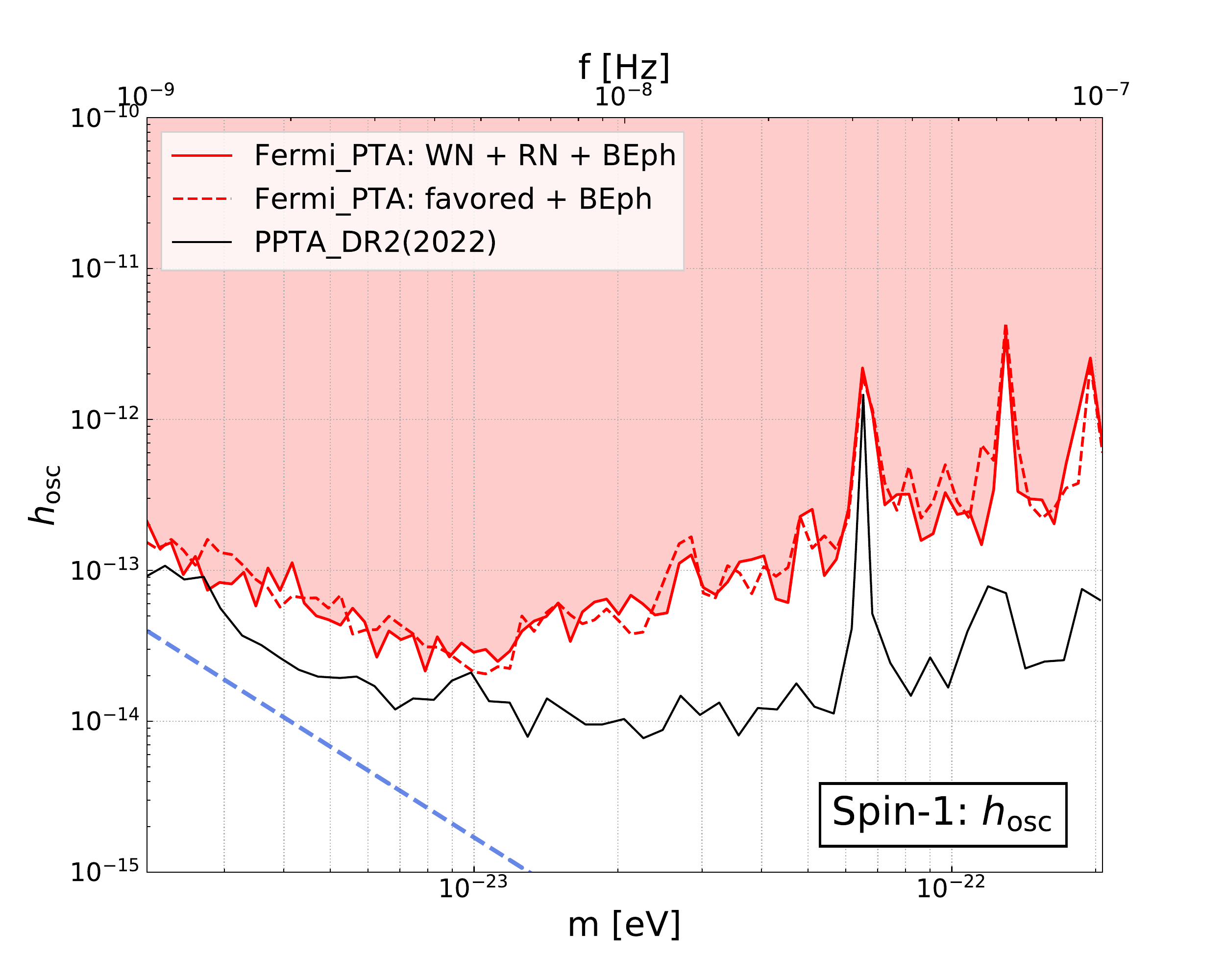}
\caption{The 95\% confidence level upper limits on oscillation amplitude of the Spin-0 (left panel) and Spin-1 (right panel) ULDM with the gravitational effect.} 
\label{Fig:upli_amplitude} 
\end{figure*}

\subsection{29 millisecond pulsars in the Fermi-LAT PTA}

Tab.~\ref{tab:29psr} lists the basic information for 29 millisecond pulsars in the Fermi-LAT PTA usd in our work: Cadence is the cadence adopted in the analysis; $d_p$ is the best estimate of the pulsar distance using the YMW16 DM-based distance taken from the ATNF Pulsar Catalogue~\cite{2005ATNF}; $\sigma_{\rm TOA}$ is the median TOA uncertainty for the Fermi-LAT PTA data; rms is the weighted root-mean-square of timing residuals; Noise Model shows the favored noise model for each pulsar taken from the Ref.~\cite{2204.05226}. The distance of PSR J2034$+$3632 is missing and set to be 1 kpc as default.

\subsection{Noise model}
The white noise of the Fermi-LAT PTA is the TOA bias, which could be larger than measured TOA uncertainties.
For each pulsar, the white noise can be added by modifying the measurement uncertainty, $\sigma$, of TOA:
\begin{equation}\label{WN}
\sigma_s^2 = ({\tt EFAC} \times \sigma)^2 + {\tt EQUAD}^2,
\end{equation}
where {\tt EFAC} is a re-scaling factor and {\tt EQUAD} is an extra uncertainty.
Ref.~\cite{2204.05226} has made a single-pulsar noise analysis for all 29 pulsars and found that only PSR J1959+2048 and PSR J2241-5236 show a preference for a model with the excess white noise.
The red noise of the Fermi-LAT PTA is mainly contributed by the intrinsic pulsar Spin-down power (called the spin noise), which can be adequately modeled with a power-law spectrum:
\begin{equation}
P(f)=\frac{A^2}{12\pi^2}\left(\frac{f}{\rm yr^{-1}}\right)^{-\gamma}~{\rm yr}^3,
\label{powerlaw}
\end{equation}
where $A$ and $\gamma$ represent the amplitude and spectral index of the power-law power spectral density, respectively.
For the radio PTA, there are other components contributing to the red noise: dispersion measure (DM) variations induced by the variable ionized interstellar
medium (IISM) in the line of sight. But for the Fermi-LAT PTA, the gamma-ray timing is immune to the turbulent IISM. 
According to the result of Ref.~\cite{2204.05226}, no evidence for red noise is found in each of all 29 pulsars.
The BayesEphem noise accounts for the systematic uncertainty of the DE421 Solar System ephemeris (SSE), which is adopted in this work, with a Bayesian modeling technique~\cite{1801.02617}.
There are 11 parameters for the BayesEphem noise: the drift-rate parameter for the semi-major axis of Earth-Moon barycenter orbit, 4 perturbation parameters for the masses of outer planets, and 6 parameters for principal components parameters of Jupiter’s orbit.
Prior distributions of noise model parameters are summarized in Tab.~\ref{tab:parameter}.

\subsection{Upper limits of oscillation amplitude of the Spin-0/1 ULDM with the gravitational effect}\label{sec:FermiPTA}

Fig.~\ref{Fig:upli_amplitude} shows constraints on oscillation amplitudes of the Spin-0/1 ULDM with the gravitational effect.
The left panel shows the 95\% confidence level upper limits of oscillation amplitude $\Psi$ for the Spin-0 case. The right panel shows upper limits of oscillation amplitude $h_{\rm osc}$ for the Spin-1 case.
For both panels, red solid lines are upper limits with all noise (WN $+$ RN $+$ BEph) components, while red dashed lines are for the case only including the favored noise model and the BayesEphem uncertainty (favored $+$ BEph).
The blue dashed lines are on behalf of the case in which the local dark matter density $\rho_{\text{DM}}$ equals to the current measurement $\rho_{0}=0.4$~GeV~cm$^{-3}$.
The black lines in two panels represent previous upper limits obtained by PPTA in Ref.~\cite{1810.03227,2210.03880}, respectively.

\bibliographystyle{apsrev4-1-lyf}
\bibliography{references}

\clearpage

\widetext

\end{document}